# NEW APPROACHES IN MODELING BELT-FLESH-PELVIS INTERACTION USING OBESE GHBMC MODELS

**Zhaonan Sun**
**Bronislaw Gepner**
**Jason Kerrigan**

University of Virginia Center for Applied Biomechanics
United States of America

## ABSTRACT

Obesity is associated with higher fatality risk and altered distribution of occupant injuries in automotive collisions partially because of the increased depth of abdominal soft tissue, which results in limited and/or delayed engagement of the lap belt with the pelvis and increases the risk of pelvis "submarining" under the lap belt exposing occupant's abdomen to belt loading. Previous modeling studies have shown that pelvis submarining could not be replicated using existing human body models. The goal of this study is to perform model modifications and investigate whether they could lead to model submarining. By detaching the connections between the pelvis and surrounding flesh, submarining like belt kinematics were observed. By remeshing the flesh parts of the model, similar belt kinematics was observed but the pelvic wings were fractured. Finally, large shear deformation on the flesh together with submarining like kinematics were observed in the model with its flesh modelled using the meshless Smooth Particle Galerkin Method (SPG) method. The results of this study showed that SPG method has potential to simulate large deformations in soft tissue which may be necessary to improve the biofidelity of belt/pelvis interaction.

## INTRODUCTION

Obesity is associated with increased fatality risk and altered distribution of occupant injuries relative to lower BMI occupants in automotive collisions [1-4]. This is partially because of the substantial effect that obesity has on occupant-restraint interaction. Restraining obese occupants is a challenge due to increased body mass, unfavorable belt placement [5], and increased forward excursion within the occupant compartment [6-7]. An increased depth of abdominal soft tissue, results in delayed and limited engagement of the lap belt with the pelvis and increases the risk of pelvis submarining under lap belt, exposing occupant's abdomen to belt loading [7].

Several experimental studies have been performed to study the challenges obesity poses on restraint system during motor vehicle collisions (MVC). Forman et al. and Kent et al. [9-10] performed rear-seat sled tests on both obese male post mortem human surrogates (PMHSs) and 50[th] percentile PMHS. It was found that obese occupants exhibited backward torso rotation (pelvis forward of shoulders) at the time of maximum forward excursion, whereas non obese occupants did not. Kent et al. [10] pointed out that obese PMHS in frontal-impact sled tests exhibited submarining behavior, which is defined as the properly-placed belt translated superiorly until it passed over the iliac crests and loaded the abdomen without engaging, or after disengaging, the pelvis [8]. The authors suggested that submarining resulted in increased forward excursion and decreased forward torso pitch, which may be related to increased risk of lower extremity and thoracic injuries in obese occupants.

HBMs with varied stature, age and BMI levels were generated using University of Michigan Transportation Institute's (UMTRI's) rapid mesh morphing tools based on statistical models of external body contour and ribcage geometry. Obese version of both Total Human Model for Safety (THUMS) [11-12], and Global Human Body Models Consortium (GHBMC) [13] were generated. While these HBMs are available to study occupant submarining, simulations illustrating HBM responses similar to PMHS kinematics are not available in the literature. A series of obese THUMS simulation were performed in [14]. Greater forward excursion was observed in frontal impact simulation, but no submarining was observed. Similar efforts have been carried out by Gepner et al. [8] with the obese GHBMC models. In the belt pull test simulations, the model did not exhibit submarining behavior as observed in the PMHS test [19]. Also, the lap belt pull simulations failed to reproduce the belt/abdomen interaction seen in the PMHS. It was also found that the material model used to represent the human body model flesh was found to be approximately one order of magnitude stiffer than human abdominal subcutaneous adipose tissue. This study shows that improved modeling of the belt-flesh-pelvis interaction should be required to obtain biofidelic response.

Experiments have suggested that adipose tissue is able to undergo substantially large shear deformations [15]. Such large deformations challenge lagrangian finite element approaches to modeling since such large deformations can result in instabilities and collapsing elements. Previous work has shown the potential to model these large deformations using meshfree methods, which offer advantages in simulating large deformation over conventional finite element methods [18]. The earliest developed meshfree method is the smoothed particle hydrodynamics (SPH) method [17]. However, this method suffers from tension instability, lack of consistency and other numerical artifacts if it is applied to solid analyses directly [16,17]. Recently, a robust and accurate meshfree method was developed by LSTC, referred to as Smoothed Particle Galerkin (SPG) method. The formulation is derived based on a smoothed displacement field within the meshfree Galerkin variational framework. It could provide stable and accurate solution for solid mechanics problems [17]. This method has been applied to manufacturing problems including drilling and metal milling. However, no previous study using this method for biological material can be found.

Since improved modeling of belt-flesh-pelvis interaction in existing HBMs are needed, and previous works showed potential in new method for large deformation modelling, the goal of this study is to apply new approaches to model belt-flesh-pelvis interaction to the obese HBMs and evaluate their ability to replicate submarining in the belt pull test.

## METHODS

The obese GHBMC models, developed by Hu et al. [13] were used as the baseline model for evaluation and modifications. Model responses were compared with the obese PMHS in a series of tabletop belt pull tests previously performed at the UVA-CAB [19]. Specifically, the PMHS with a BMI of 31 kg/m2, height of 1650 mm and weight of 84.4 kg was chosen for comparison with the GHBMC obese model with a height of 1750 mm and a BMI of 30 kg/m2. FE model of the test fixtures used in the belt pull tests previously developed using 3D drawings of the original test fixture were used in this study Gepner et al. [8]. The baseline belt position, prescribed force time history was adapted from Gepner et al. [8], which was based on the experiment. Three independent methods for model modification were applied. First, the connection between the pelvis and surrounding tissue was removed, enabling relative motion between pelvis and flesh. Second, spatial discretization of the abdominal flesh was improved to obtain better resolution and sensitivity study on material stiffness were performed on the remeshed model. Third, SPG method was introduced to model the outer flesh to enable large deformation.

**Contact modification**

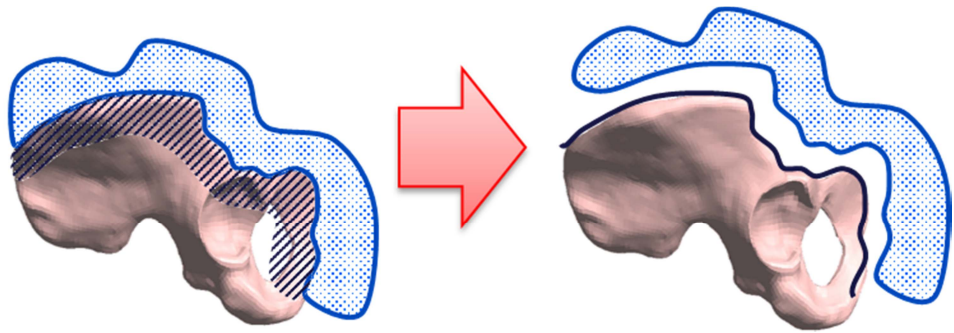

**Figure 1 Schematic of detaching the surrounding soft tissue from the pelvis**

In the obese GHBMC models, the pelvic wings are connected to surrounding tissue through tied contact and shared nodes (Figure 2). This could potentially prevent the large shear motion within the elements of surrounding soft tissue since some element always have to be connected to the bony pelvis. Since we theorized that the key to

Sun 2

submarining is the large shear motion on the flesh, we released the constraints between the pelvis and the surrounding flesh and explore whether this would enable large shear motion. The schematic of this detachment can be seen in Figure 1. Specifically, we released the tied contacts and shared nodes relation between the parts and ran the belt pull test simulation with three different input pulses. The first pulse is the baseline pulse fitted from the PMHS test. The second pulse was obtained by increasing the peak force to compensate for difference in anthropometry between the HBM and PMHS. Finally, by increasing the holding time of the peak force, the overall energy input was increased to obtain the third pulse.

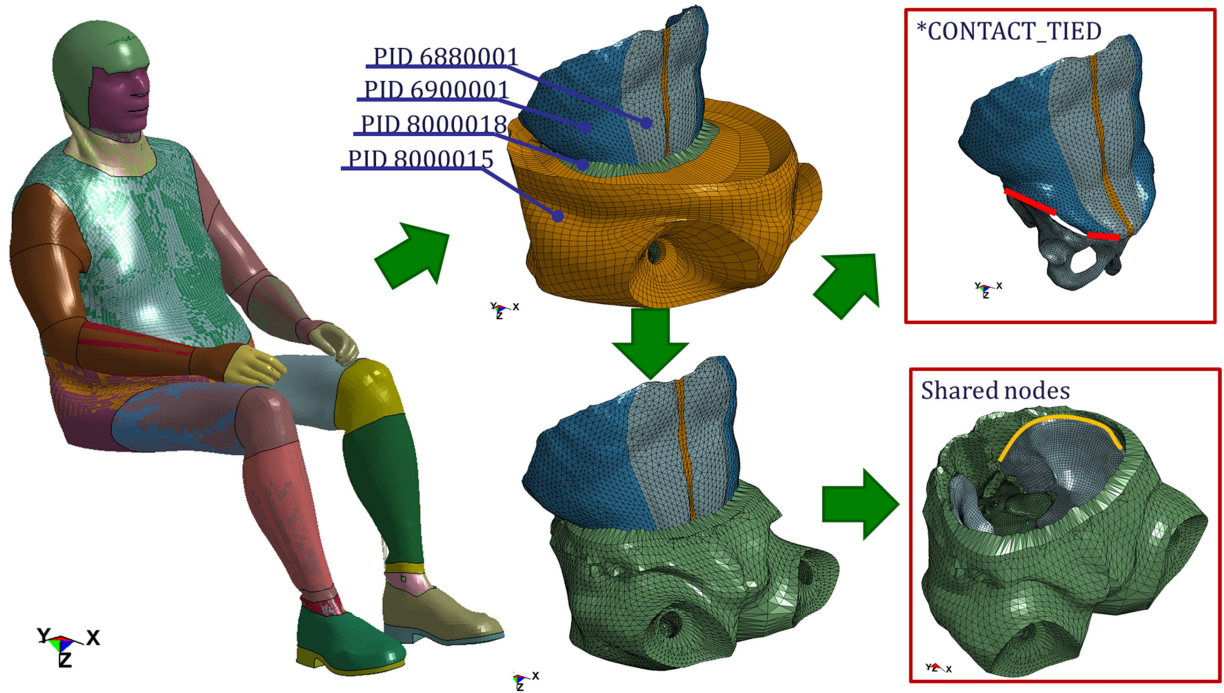

**Figure 2 Connections between the pelvis and surrounding tissue in the obese GHBMC model**

**Model Remeshing**

The obese GHBMC models were generated with rapid mesh morphing tools based on statistical models of external body contour and ribcage geometry. During this process, only nodal coordinates changed. Contact definition, material properties and element definition remained unchanged. To account for differences in external body contour of an obese occupant, abdominal and thoracic flesh layer were thickened by stretching the elements in the morphing process. As shown in Figure 8, the obese GHBMC only used three elements across the abdominal front wall. Since these elements were directly stretched from elements in the $50^{th}$ percentile GHBMC model, they all have suboptimal aspect ratios. To improve mesh quality, remeshing was performed on two solid and two shell parts. Both solid parts were meshed with constant density tetrahedral elements. The number of elements increased by around 50,000 after the remeshing. A baseline simulation was first performed, followed by a three times baseline simulation. Then, pelvic wing yielding criteria was turned off and a three times baseline simulation was performed. Finally, material property was adjusted, and a baseline simulation was performed.

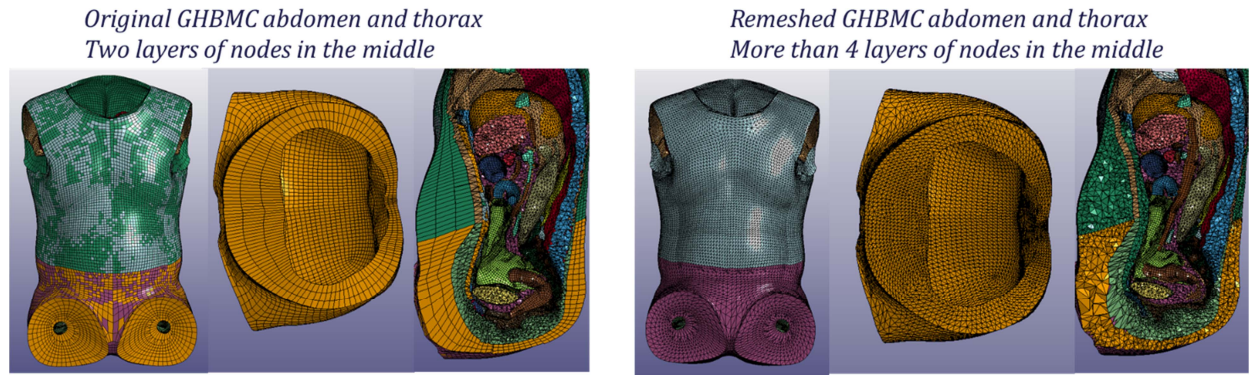

**Figure 3 Remeshing the abdominal and thoracic flesh (Complete models with head, upper and lower extremities hidden )**

**SPG implementation**
Although SPG is a meshless method by nature, the creation of SPG particle is mesh based. After creating the mesh, LS-DYNA solver turns each node in the selected SPG section into one smooth particle and discard the solid element connection between the particles. Since the method of SPG performs well with at least 3 layers of smoothed particles in between the lagrangian boundaries and particle needs to be distributed as evenly as possible, the remeshed model nodal coordinates were used to turn the outer thoracic and abdominal flesh nodes into smoothed particles while the skin and inner flesh remained in lagrangian formulation. More than 10 layers of SPG particles were generated at every nodal position in between the lagrangian boundaries of the remeshed model but the nodal connectivity within the flesh parts were neglected in the computation. In this study, two simulations were performed with three times of baseline input. One of them has pelvic wing yielding criteria turned off while the other has it remained on. Another simulation was performed with 1.5 times baseline input and with a less stiff flesh material model.

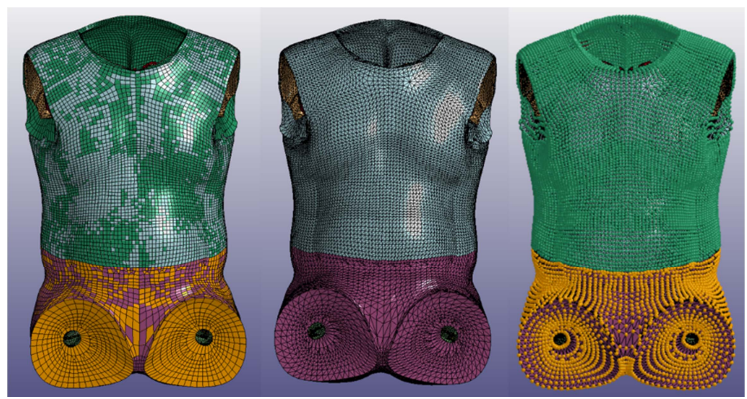

**Figure 4 SPG particles generated from remeshed flesh parts**

## RESULTS

For all belt pull simulations, graphs showing the comparison of belt trajectories obtained from the experiment and the simulation, as well as time histories of belt pulling force and belt displacement are presented in the results session. For the belt trajectories, the time histories for points on the belt directly anterior to ASISL and ASISR were extracted and plotted in green dots over the experimental results in blue dots.

**Detached pelvis to flesh**
When the detached model was pulled with the baseline force that fitted to the time history from the experiment, no submarining was observed (Figure 5). The blue dots showed that in the experiment, the belt started moving towards the ASIS and then changed direction to load the abdomen. The green dots showed that, in the simulation, the belt went straight towards the pelvis without showing trend for direction change.

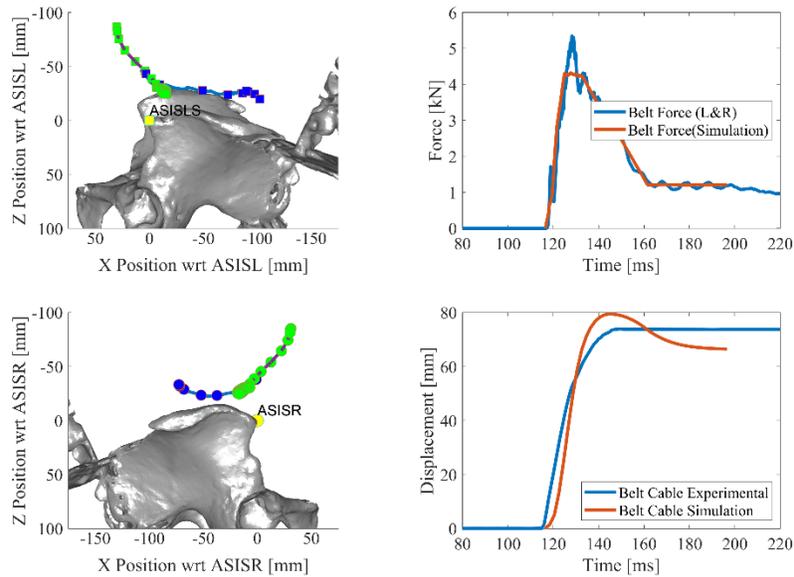

**Figure 5 Detached pelvis with a fit-to-experiment input pulse**

When the amount of energy input to the system was increased by both increasing the peak force and its holding time before release, highly matched belt trajectory with the experiment was observed (Figure 6). The belt first compressed the abdomen towards the pelvis, then under the influence of the reaction force from the pelvic wings, changed its direction and moved in to compress the abdominal cavity.

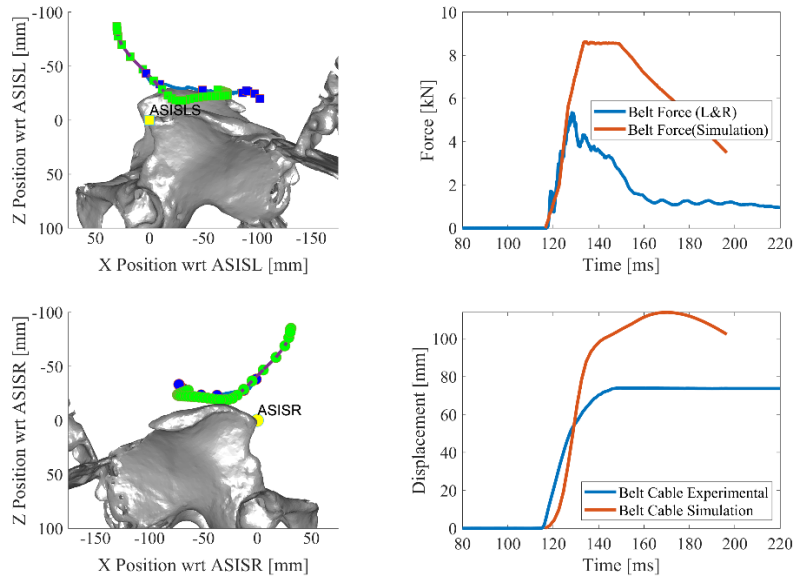

**Figure 6 Detached pelvis with added energy double peak input**

**Remeshed abdomen and thorax**

Under baseline loading conditions, the model could not submarine due to lack of energy (Figure 7). When peak force was increased to three times of baseline, the model submarined (Figure 8). However, the pelvic wing elements reached their yielding point and started to fail. By turning off the failure criteria on the

Sun 5

pelvic wings, the same input condition led to a negative volume problem in the flesh and no submarining was observed (Figure 9).

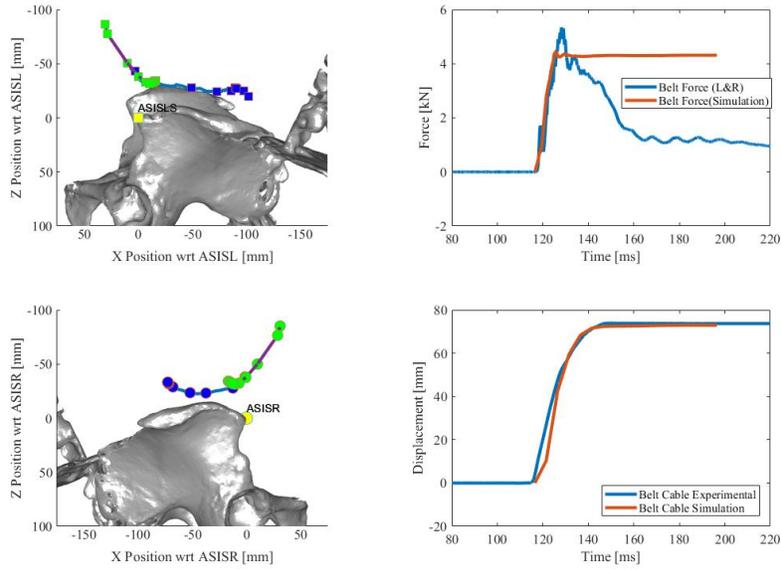

**Figure 7 Baseline flat input simulation in lagrangian formulation with complete boundary condition**

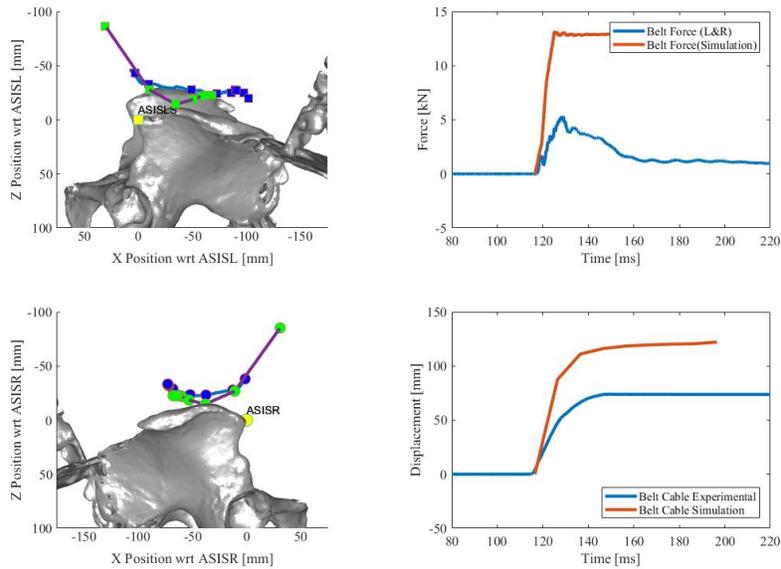

**Figure 8 Three times peak force and rate flat input simulation in lagrangian formulation with complete boundary condition**

Sun 6

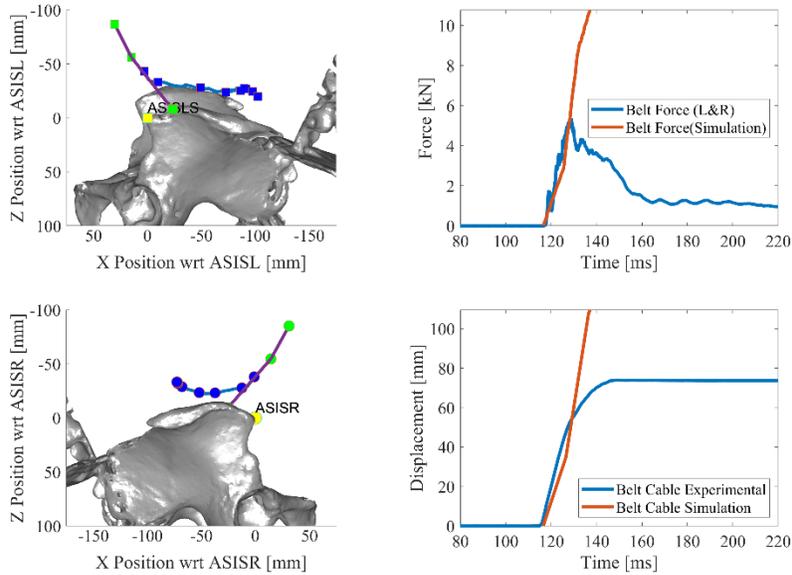

**Figure 9 Three times peak force and rate flat input simulation in lagrangian formulation with pelvis failure turned off**

### SPG abdomen and thorax

Similar to the remeshed lagrangian model, under baseline loading conditions, the SPG model did not submarine possibly due to lack of energy. With a three times baseline force input, the model submarined but the pelvic wing elements were failed when the pelvis failure criteria was on. When turning off the pelvic failure criteria, the belt managed to load the abdomen into large shear without running into negative volume problem as encountered in the lagrangian model. A comparison revealed that lap belt found a different path to load the abdomen, avoiding loading the pelvic wings to failure (Figure 10).

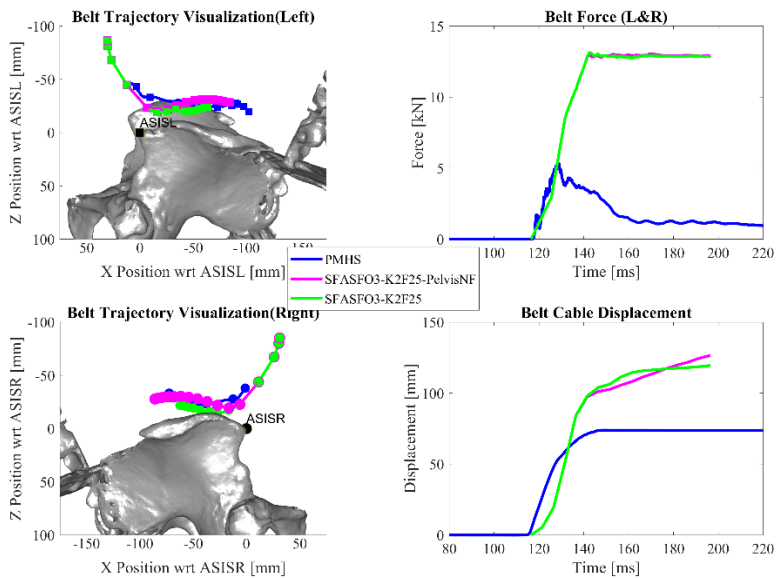

**Figure 10 Comparison between both three times peak force and baseline rate flat input simulation in SPG formulation with (in green) or without pelvis failure (in pink)**

## DISCUSSIONS

**Detached pelvis to flesh**

When pulling the detached pelvis model with the baseline force, the model did not submarine due to insufficient energy. First, the peak force was hold for less than 20ms, leaving not enough time for belt movement. Second, the obese GHBMC in the simulation is an overall larger person than the PMHS, therefore requiring more energy to achieve the same amount of deformation. Third, [8] we concluded that abdominal flesh in the Obese GHBMC appeared to be orders of magnitude stiffer in shear compared to subcutaneous adipose tissue. These three reasons would affect the pulling force required to reach desired deformation, therefore lead to the non-submarining results.

When the input force was scaled to two times the baseline and held longer, the global belt kinematics matched with the belt pull test well. We concluded this type of simulation as the type I submarining. However, type I submarining is not what happens in human body since surrounding muscles are attached to the pelvic wings. Computationally, the original boundary condition of the belt pull test was broken actively by detaching surrounding tissue from the pelvis. As a result, this method should not be used to modify the current obese GHBMC models for further simulations.

**Remeshed abdomen and thorax**

The remeshed model submarined since pelvic wing being sheared off to failure broke the original boundary conditions. We define this as the type II submarining, also the pseudo submarining. Turning off the pelvic wing failure criteria would prevent pseudo submarining from happening. This can be achieved in two ways. The first method is to increase the yielding point to make it almost not reachable in the current simulation setting. The second method is to turn off the yielding criteria directly. Both methods gave the same response. After turning off the failure criteria on pelvic wings using either method, negative volume in flesh happened. This showed that by keeping the boundary condition intact, the remeshed model could not submarine.

**SPG abdomen and thorax**

To the best of authors' knowledge, this is the first simulation study using SPG to model soft tissue in HBMs. This new modeling approach resulted in model submarining in the belt pull test simulation. Gepner et al. [8] theorized that, in the belt pull test, the belt is initially pulled towards the pelvis and then encounters the bony pelvis, which provides a reaction force to guide the belt over the iliac crest and into the abdomen. In this study, when the abdominal flesh is modelled with SPG, tissue compression first happened. Then, under the guidance of the reaction force from pelvic wings, the lap belt changed its direction and went over the iliac wings. By applying proper parameters for the SPG particles, it was found that belt displacement over the pelvic wings can be controlled. Also, by turning off the pelvic wing failure parameter, the belt managed to navigate a different path to submarining, which appears to be superior to the one with pelvic wing failure parameter. This difference in trajectory showed that turning off the pelvic wing failure criteria did change the overall kinematics of the belt.

It is worth noting that this is not a model validation study since the HBM in this study has different anthropometry from the PMHS being compared. However, this study showed that three types of model modifications can make the model reach desired kinematics. Specifically, this study supported the theorized mechanism of submarining by Gepner et. al [8] by showing promising results in the SPG models. However, potential contact and stability issues still exists and need to be further evaluated.

## CONCLUSIONS

The following conclusions can be drawn from this study:

1. Submarining could happen due to breaking of boundary condition or large shear deformation in flesh.

2. Detaching the connection between the pelvis and surrounding flesh could break the boundary condition and therefore lead to submarining.

3. Pulling hard with a denser mesh in the flesh could break the boundary condition and therefore lead to submarining.

4. Large shear deformation can be realized through using SPG particle to particle bond failure criteria.

5. Tuned SPG parameters worked well in the belt pull test simulation, recreating similar kinematics in the Obese GHBMC model to the PMHS